\documentclass[12pt]{article}
\makeatletter
\def\section{\@startsection {section}{1}{\z@}{-3.5ex plus -1ex minus 
 -.2ex}{2.3ex plus .2ex}{\large\bf}} 
\def\subsection{\@startsection{subsection}{2}{\z@}{-3.25ex plus -1ex  
minus -.2ex}{1.5ex plus .2ex}{\normalsize\bf}}

\newcommand{\ket}[1]{|{#1}\rangle}

\def\beq{\begin{equation}}  
\def\eeq{\end{equation}}  
\def\beqa{\begin{eqnarray}}  
\def\eeqa{\end{eqnarray}}  
\newcommand{\sect}[1]{\setcounter{equation}{0}\section{#1}}  
\renewcommand{\theequation}{\thesection.\arabic{equation}}  
\newcommand{\EQ}{\begin{equation}}  
\newcommand{\EN}{\end{equation}}  
\newcommand{\bea}{\begin{eqnarray}}  
\newcommand{\ena}{\end{eqnarray}}  
  
\renewcommand{\a}{\alpha}

\newcommand{\dpb}{D$p$-brane}  
\newcommand{\dpbs}{D$p$-branes}  
\newcommand{\db}{D$0$-brane}  
  
\renewcommand{\thefootnote}{\fnsymbol{footnote}}  
\def\one{{\hbox{ 1\kern-.8mm l}}}

\newlength{\bredde}  
\def\slash#1{\settowidth{\bredde}{$#1$}\ifmmode\,\raisebox{.15ex}{/}  
\hspace*{-\bredde} #1\else$\,\raisebox{.15ex}{/}\hspace*{-\bredde} #1$\fi}  
  
\renewcommand{\thefootnote}{\arabic{footnote}}  
\begin{document}  
\begin{titlepage}  
\rightline{DFTT 42/2000} 
\rightline{DSF 35/2000} 
\vskip 1.8cm  
\centerline{\Large \bf THE GEOMETRY OF FRACTIONAL BRANES  
\footnote{Work   
partially supported by the European Commission  
RTN programme HPRN-CT-2000-00131 in which R.M. is associated with 
Frascati-LNF, and by MURST.}} 
\vskip 0.8cm  
\centerline{\bf M. Frau$^a$,  
A. Liccardo$^{a}$ 
R. Musto $^b$  
\footnote{R.M. dedicates this  
work to the memory of Lochlain O' Raifeartaigh.}}   
\vskip .5cm  
\centerline{\sl $^a$ Dipartimento di Fisica Teorica, Universit\`a di  
Torino}   
\centerline{\sl and I.N.F.N., Sezione di Torino, Via P. Giuria 1, I-10125   
Torino, Italy}  
\centerline{e-mail: surname@to.infn.it} 
\vskip .2cm  
\centerline{\sl $^b$ Dipartimento di Scienze Fisiche, Universit\`a di Napoli}  
\centerline{\sl and I.N.F.N., Sezione di Napoli,   
Monte S. Angelo, I-80126 Napoli, Italy}  
\centerline{e-mail: surname@na.infn.it} 
\vskip 1.3cm  
\begin{abstract} 
 
By looking at fractional {\dpbs} of type IIA on  
${\bf T}_4/{\bf Z}_2$ 
as wrapped branes 
and by using boundary state techniques we construct the effective low-energy 
action for the fields generated by 
fractional branes, build their world-volume action and find the  
corresponding classical geometry. 
The explicit form of the classical background is consistent only outside  
an  enhan\c{c}on sphere of radius $r_e$, which encloses a 
naked singularity of  
repulson-type. 
The perturbative running of the gauge coupling constant, dictated by the  
NS-NS twisted field that keeps its one-loop expression at any distance,  
also fails at $r_e$.  
 
\end{abstract}  
\end{titlepage}  
\newpage  
\tableofcontents  
\vskip 1.5cm 
\renewcommand{\thefootnote}{\arabic{footnote}}  
\setcounter{footnote}{0}  
\setcounter{page}{1}   
\sect{Introduction}  
\label{introduction}  
\vskip 0.5cm  
The duality relation between string and gauge theories \cite{mal}   
has been studied in detail in the AdS/CFT case, but recently more   
effort has been devoted to study the gauge/gravity correspondence  
in the case of non conformal gauge theories and Yang-Mills theories  
with reduced supersymmetry \cite{MILANO}-\cite{raja}.  
Fractional branes \cite{FRACTIONAL}, that arise for example in orbifold   
models, provide a natural set-up to probe the Maldacena conjecture when  
supersymmetry is partially broken with respect to the $\cal N$=4 case  
and conformal invariance is lost.  
The full knowledge of space-time geometry of fractional branes is then not  
only an interesting problem relative to singular geometries and their  
resolution but also a physical issue providing a new insight for  
understanding the stringy aspects of gauge theory in   
more realistic models.  
  
In this paper we will discuss the case of fractional {\dpbs} of type IIA  
string theory compactified on $T_4/ Z_2$,  
where $Z_2$ is generated by the parity operator on the four compact 
spatial coordinates.  
While the techniques we use to reconstruct the space-time geometry 
associated to fractional branes are rather general, for the sake of  
simplicity we will mainly refer to the case of a fractional {\db} 
of type IIA theory on $T_4/ Z_2$.   
As it is well known, this is an orbifold limit of the type IIA string 
compactified on the smooth hyperK\"ahler manifold K3.  
Within this frame a fractional D$p$-brane can be interpreted as  
a D$(p+2)$-brane wrapped on one of the supersymmetric two-cycles of K3 that  
vanish in the orbifold limit.  
This point of view provides a natural {\it ansatz}   
to rewrite the relevant massless fields of the ten dimensional IIA  
orbifold theory in terms of fields coupled to the brane. Then one obtains 
the truncation of   
the low energy action of IIA on $T_4/Z_2$ that provides the dynamics of  
fields coupled to fractional {\dpbs}.  
In order to reconstruct the full geometry of a fractional brane one  
also needs the expression of relevant source terms, namely of a  
boundary action. This information can be obtained by analyzing the  
structure of the boundary states \cite{bs}.  
The use of the boundary state technique \cite{antonella} is rather   
natural in this context as it translates open string boundary conditions  
in terms of closed string, introducing automatically into the game gravity  
together with all other fields interacting with the fractional brane.  
The analysis of the couplings between the massless fields and the boundary  
state allows us to infer the structure of the complete world-volume action   
of fractional branes on the compact orbifold, which turns out to be  
also an essential tool for analyzing the motion of a probe brane  
in the classical background. 
 
One expects that  the space-time geometry of fractional branes shares a  
common feature with other classical backgrounds which are dual to non  
conformal gauge theories, namely the presence of naked singularities of  
{\it repulson} type.  
This feature has been found 
in the study of fractional branes on singular spaces 
\cite{JPP,KLEBA1,KLEBA2,KLEBA3,TSEYTLIN1},  
of stable non-BPS branes \cite{BERT1} and of type IIB 
fractional branes on $R_{1,5}\times R_4/Z_2$ \cite{BERT2,POL2}.  
Our detailed investigation on the D$0$-brane solution and its generalization 
to the D$2$-brane, shows that this is the case also for fractional branes  
of type IIA on $T_4/Z_2$ .   
However, also in our case, like in 
\cite{JPP,BPP,PETRINI}  this   
singularity  is  actually unphysical as it is  outside the  
region where the simple supergravity approximation is reliable.  
In fact, at a distance  $r_e$ greater   
than the one where the singularity is located, a probe fractional brane  
becomes tensionless, and at $r=r_e$ there is a geometric locus, 
called {\it enhan\c{c}on}, 
where extra massless degrees of freedom become relevant 
(for a review on the probe technique see {\it e.g.} Ref.~\cite{CVJ}).  
In this case, as discussed in Ref. \cite{JPP}, the fractional branes building  
up the classical background are forced to cover uniformly the hypersphere  
at $r=r_e$ rather than pile-up at $r=0$.  
Then at a distance shorter than the enhan\c{c}on radius   
the supergravity description is not valid any more   
and one has to modify the effective theory.  
  
In this paper we do not study in detail the connection between the  
fractional brane classical solution and the world volume gauge  
theory and we leave it to a future work.  
We just mention that in general the running of the coupling constant of  
the dual gauge theory turns out to be dictated by the behavior of the  
twisted fields and in particular of the NS-NS twisted field.  
A remarkable property of our solution, is that   
twisted fields keep their harmonic asymptotic form at any distance.   
This appears as a general feature for fractional branes,  
apparently translating in geometrical classical terms the ${\cal N}=2$ 
SUSY gauge  
theory property of allowing only one-loop perturbative corrections.    
This is in fact the case for type IIB fractional D3-brane 
\cite{BERT2,POL2}, where the appropriate perturbative 
logarithmic behavior of  
Yang-Mills coupling constant   
in terms of a stringy description has been obtained.  
  
The paper is organized as follows: section {\bf 2} is devoted to the   
reduction of the low energy effective action for type IIA theory in   
ten dimension to the six dimensional action describing the dynamics  
of the fields coupled to a fractional D$p$-brane on $T_4/Z_2$.   
The boundary state technique, which is used to obtain   
a boundary action and the asymptotic behavior of relevant fields, is   
described in some detail in section {\bf 3}.   
In section {\bf 4} we give the solution of the   
equations of motion and discuss their physical implications. The three   
appendices are devoted respectively to an alternative derivation of  
the low energy action based on the $S$-duality between   
IIA on $T_4/Z_2$ and heterotic on $T_4$ (appendix {\bf A}),   
to the analysis of the massless spectrum   
of the $Z_2$ orbifold theory (appendix {\bf B}) and to   
the explicit derivation of the equations of motion and their   
solution (appendix {\bf C}).      

\vskip 1.5cm 
\sect{The low energy effective action} 
\label{action} 
\vskip 0.5cm 
One of the ingredients necessary to determine the geometry of a  
fractional brane is the low energy action expressing the dynamics of the 
relevant fields in the bulk. 
In this section we start with type IIA supergravity in ten dimensions  
with a ${ Z}_2$ orbifold projection acting as a reflection on four 
coordinates  
(which we choose to be $x_6,x_7,x_8,x_9$). 
Introducing a specific {\it ansatz} for the fields, we obtain the peculiar  
truncation of the low energy action which is appropriate to describe  
the fractional brane in 
$R^{1,5}\times R^{4}/ Z_2$. 
Then, by performing a Kaluza-Klein reduction to six dimensions,  
we get the low energy action for fractional branes in the orbifold  
$R^{1,5}\times T_{4}/Z_2$. 
In appendix {\bf A} we obtain the same action exploiting the $S$-duality  
relation between our theory and  
heterotic theory compactified on $T_4$. 
 
The ten dimensional type IIA supergravity action on the orbifold 
$R^{1,5}\times R^{4}/Z_2$ 
in the string frame can be written as 
\[ 
S=\frac{1}{2\kappa^2} \left\{\int d^{10}x\sqrt{-G}\,e^{-2\Phi}\,R(G)+ 
\int e^{-2\Phi}\left[4\; d\Phi\wedge^*d\Phi-\frac{1}{2}H_3\wedge^*H_3\right] 
\right. 
\] 
\beq 
\label{ac1} 
\left. 
+\frac{1}{2}\left[F_2\wedge^*F_2 + \widetilde F_4\wedge^*\widetilde F_4 
-B_2\wedge F_4\wedge F_4\right] 
\right\} 
\eeq 
where 
\beq 
\label{ac2} 
H_3=dB_2\,\,\,\,,\,\,\,\, 
F_2=d{\widetilde C}_1\,\,\,\,,\,\,\,\,F_4=dC_3 
\eeq 
are respectively the field strengths of the Kalb-Ramond field, 
of the R-R vector field and of the R-R 3-form potential 
and  
\beq 
\label{ac3}  
\widetilde F_4=F_4-{\widetilde C}_1\wedge H_3~~. 
\eeq 
Moreover $\Phi$ is the 10-dimensional dilaton fluctuation and 
$\kappa =(2\pi)^{7/2}\alpha'^2 g_S$ is the gravitational coupling 
constant appropriate to our orbifold background. 
 
A fractional D$p$-brane can be interpreted as a D$(p+2)$-brane 
wrapped on the vanishing 2-cycle of the orbifold \cite{ASPINWALL}. 
According to this interpretation we make the following {\it ansatz} 
on the Kalb-Ramond  and R-R 3-form potential  
\beq 
\label{ac4} 
B=b\,\omega_2\,\,\,\,\,\,,\,\,\,\,\,\,{C}_3=A_1\wedge\omega_2 
\eeq 
where $\omega_2$ is the closed 2-form dual to the vanishing 2-cycle 
which we normalize in such a way that  
\beq 
\label{ac7} 
\int\omega_2\wedge^*\omega_2=1~~, 
\eeq  
and the scalar field $b$ is  
\beq 
\label{ac4b} 
b=\frac{1}{2}(2 \pi \sqrt{\alpha '})^2 + D ~~. 
\eeq 
 
Using the previous equations  
in Eq.(\ref{ac1}) we obtain the following expression for the action  
\[ 
S =\frac{1}{2\kappa^2} 
\Bigg\{\int d^{10}x\sqrt{-G}\,e^{-2\Phi}\,R(G)+ 
\int \left[4\,e^{-2\Phi}\; d\Phi\wedge^*d\Phi+\frac{1}{2}F_2\wedge^*F_2\right] 
\] 
\beq 
\label{ac6}
-\frac{1}{2}\int_6\left[e^{-2\Phi} dD\wedge^*dD-
(dA_1-{\widetilde C}_1\wedge dD) \wedge^* 
(dA_1-{\widetilde C}_1\wedge dD)\right] 
\Bigg\} 
\eeq 
where the index $6$ means that the fields are integrated over the  
$6$-dimensional space-time  which is unaffected by the orbifold projection. 
Notice that under the {\it ansatz} (\ref{ac4}) the Chern-Simon term  
appearing in eq. (\ref{ac1}) vanishes, showing that it does not contribute  
to the dynamics of fractional branes. 
The action appropriate to the case of the compact orbifold $T_4/Z_2$ 
can now be obtained from (\ref{ac6}), by making a simple Kaluza-Klein  
reduction of the ten dimensional part 
\[ 
S=\frac{\cal V}{2\kappa^2}\Bigg\{ 
\int d^{6}x\sqrt{-g}e^{-2\varphi}R(g)+ 
\int_6 e^{-2\varphi}\left[ 
4 d\varphi\wedge^*d\varphi 
-d{\widetilde\eta_{a}}\wedge^*d{\widetilde\eta^{a}} 
\right]
\] 
\[ 
\left. 
+\frac{1}{2}\int F_{2}\wedge^* 
F_{2}\prod_a e^{\widetilde\eta_a} 
\right\} 
-\frac{1}{2\kappa^2} 
\int_6 
\left\{\frac{e^{-2\varphi}}{2}\prod_a e^{-\widetilde\eta_a}  
dD \wedge^*dD+ 
\right. 
\] 
\beq 
\label{ac8} 
-\frac{1}{2}\left[F_{2}^A\wedge^*F_{2}^A-2F_{2}^A\, 
\wedge^*({\widetilde C}_{1}\wedge dD)+ 
{\widetilde C}_{1}\wedge dD\,\wedge^*({\widetilde C}_1\wedge dD) 
\right] 
\Bigg\} 
\eeq 
where $g_{\mu\nu}$ is the six dimensional metric, while 
the 6-dimensional dilaton field $\varphi$ and 
the scalar fields $\tilde\eta_a$ are defined as follows 
\beq 
\label{ac9} 
\varphi=\Phi-\frac{1}{4}\ln\left(\prod_a G_{aa}\right)\,\,\,\,\,\,\,, 
\,\,\,\,\,\,\, 
G_{aa}=e^{2\widetilde\eta_a} 
\eeq 
with $a= 6, 7, 8,9 $. Moreover we have introduced 
$F_2^A=dA_1$ and denoted the volume of $T_4$ by  
\beq 
{\cal V}=\prod_a(2\pi R_a) ~~~~. 
\label{V} 
\eeq 
Defining  $\kappa_{\rm orb}= \kappa/{\cal V}^{1/2}$  
and going to the Einstein frame, we get 
\[ 
S=\frac{1}{2\kappa_{\rm orb}^2}\left\{ 
\int d^6 x\sqrt{-g}\,{R}(g)  
+\int_6\left[-d\varphi\wedge^*d\varphi 
-d{\widetilde\eta_{a}}\wedge^*d{\widetilde\eta^{a}}+ 
\right. 
\right. 
\] 
\[ 
\left. 
+\frac{1}{2}e^{\varphi} 
\prod_a e^{\widetilde\eta_a}{F}_2\wedge^*{F}_2 
-\prod_{a}e^{-\widetilde\eta_a}d{\tilde D}\wedge^* d{\tilde D}+ 
\right. 
\] 
\beq 
\left. 
\left. 
+\frac{e^\varphi}{2} 
\left( 
{\widetilde F}_2^A\wedge^*{\widetilde F}^{A}_2  
-2\sqrt{2} 
{\widetilde F}_{2}^A 
\wedge^*({\widetilde C}_1\wedge d{\tilde D})+ 
2{\widetilde C}_1{\wedge d\tilde D}\wedge^*  
({\widetilde C}_1\wedge d{\tilde D}) 
\right) 
\right] 
\right\} 
\label{ac10} 
\eeq 
where we have made the following rescaling  
\beq 
\label{act11}   
\widetilde A_1= \frac{A_1}{\sqrt{\cal V}}~~~~,~~~~ 
\tilde D=\frac{D}{\sqrt{2{\cal V}}}~~. 
\eeq 
The action (\ref{ac10}) describes the fields that 
couple to an electric fractional D$0$-brane,  
or equivalently to a magnetic fractional D$2$-brane. 
In next section, using the boundary state formalism, we will  
instead obtain the asymptotic behavior of the classical fields  
generated by a generic electric fractional D$p$-brane.  
In solving the classical equations of motion derived  
from (\ref{ac10}) with the boundary conditions 
dictated by the boundary state, it is then more  
natural to address the case of the D$0$ fractional brane. This is  
what we will actually do in detail in the following. 
The explicit solution found in that case can be however 
easily generalized also to case of D$2$ fractional brane, as we  
will see in section {\bf 4}. 

\vskip 1.5cm  
\sect{The boundary state description}  
\label{boundary}  
\vskip 0.5cm  
  
The boundary state formalism allows one to connect    
microscopic and  supergravity description of D-branes   
by using the language of closed string  
state to evaluate  both  the couplings of    
supergravity massless fields to the brane and their asymptotic behavior.  
The boundary state description of fractional D$p$-branes   
in type II theories on orbifold  
has been already discussed in Refs. \cite{diaco,gabstef,taka} in the case   
of ${Z}_N$ projection, and for general   
discrete groups in Ref.~\cite{BILLO}.  
Here we briefly review the structure of the boundary state in the simple   
case of ${Z}_2$, in order to derive the relevant physical   
information about fractional branes.  
  
The boundary state is a closed string state which embodies the presence   
of a {\dpb} in space-time and is defined by the overlap equations, which,  
in the notation and conventions of Ref.\cite{antonella}, read as  
\beq  
\label {bc1c}  
\partial_{\tau}X^\alpha|_{\tau=0}|B\rangle =0 ~~~~~,~~~~~\alpha =0,...,p~~,  
\end{equation}  
\begin{equation}  
\label {bc2c}  
X^i|_{\tau=0}|B\rangle =y^i|B\rangle ~~~~~,~~~~~i = p+1,..., d-1~~.  
\end{equation}  
and  
\beq  
\label {psib}  
(\psi^\mu_-(\sigma,0)-i\eta S^{\mu}_{\,\,\, \nu} {\psi^\nu_+}(\sigma,0))  
|B_{\psi}, \eta \rangle  =0  
\eeq  
where $S^{\mu\nu}=(\eta^{\alpha\beta},-\delta^{ij}),$ and  $\eta=\pm 1$.  
  
As it is well known, the spectrum of type II strings on orbifolds   
consists not only of states of the original theories which are invariant   
under the orbifold projection, the untwisted sector, but also of states   
which belong to the so-called twisted sectors living at the orbifold   
fixed hyperplanes.  
Therefore 
in an orbifold theory the complete boundary state is a linear   
combination of all the Ishibashi states coming from the resolution of   
eqs.(\ref{bc1c})-(\ref{psib}) in each sector.  
  
In the specific case of the orbifold $T_4/{Z}_2$, there are 16 distinct   
twisted sectors, one for each fixed plane. However, when the 
fractional brane   
does not wrap in the compact directions, there are only   
two different coherent states which solve the overlap equations,   
one for the untwisted and the other for the twisted sector associated to  
the particular fixed plane on which the brane lives. 
  
In this particular case we can therefore write   
\beq  
\label{bound1}  
\ket{B}= {\cal N}^U\left(\ket{B}_{\rm NS}^U+  
\epsilon_1\ket{B}_{\rm R}^U  
\right)+{\cal N}^T\left(\ket{B}_{\rm NS}^T+\epsilon_2\ket{B}_{\rm R}^T  
\right)  
\eeq  
where ${\cal N}^U$ and ${\cal N}^T$ are two normalization  
constant to be fixed,  
$U$ and $T$ stand for untwisted and twisted respectively and    
$\epsilon_1$,$\epsilon_2$ are the  
two R-R charges in the untwisted and twisted sectors.   
The states $\ket{B}^U$ are the usual boundary states of a bulk-brane on   
a compact space with the standard GSO-projection \cite{noi},  
while the states   
$\ket{B}^T$ will be   
constructed in the following subsection.  
  
Notice that the boundary state in (\ref{bound1}) is 
the one associated to the trivial 
representation of the ${Z}_2$ group on Chan-Paton factors of  
open strings attached to fractional branes \cite{BILLO}.  
  
\vskip 1cm  
\subsection{Construction of the boundary state}  
\vskip 0.5cm  

The Ishibashi states of the twisted sector, satisfying the overlap   
conditions (\ref{bc1c})-(\ref{psib}), can be written as
\beq  
\label{bound2}  
\ket{B,\eta}^T_{\rm NS}=   
\ket{B_X}^T\ket{B_\psi,\eta}_{\rm NS}^T  
\eeq  
in the NS-NS twisted sector and similarly  
\beq  
\label{bound3}  
\ket{B,\eta}^T_{\rm R}=   
\ket{B_X}^T\ket{B_\psi,\eta}_{\rm R}^T  
\eeq  
in the R-R twisted sector \footnote{In (\ref{bound2}) and (\ref{bound3})  
we omit the ghost and superghost contribution   
which is not affected by the   
orbifold projection.},  
where   
\bea  
\label{bound4}  
|B_X \rangle^T  &=&\delta^{5-p}({\hat q}^i-y^i)   
\prod_{n=1}e^{-\frac{1}{n}  
\alpha_{-n}^\mu  
S_{\mu\nu}\tilde\alpha_{-n}^\nu}  
\prod_{r=\frac{1}{2}}e^{\frac{1}{r}  
\alpha_{-r}^a\tilde\alpha_{-r}^{\,a}}  
|0\rangle _{\alpha}|0\rangle  
_{\tilde\alpha}  
|p=0\rangle   
\\  
\label{bound5}  
|B_{\psi} , \eta \rangle_{NS}^T  &=&   
-\prod_{t=\frac{1}{2}}e^{i\eta\psi_{-t}^\mu S_{\mu\nu}   
\tilde \psi_{-t}^\nu}   
\prod_{t=1}e^{-i\eta\psi_{-t}^a   
\tilde \psi_{-t}^{\,a}} |B_{\psi} , \eta \rangle ^{(0)\,\,T}_{\rm NS}  
\\  
\label{bound6}  
|B_{\psi} , \eta \rangle_{R}^T &=&  
\prod_{t=1}e^{i\eta\psi_{-t}^\mu S_{\mu\nu}   
\tilde \psi_{-t}^\nu}   
\prod_{t=\frac{1}{2}}e^{-i\eta\psi_{-t}^a    
\tilde \psi_{-t}^{\,a}} |B_{\psi} , \eta \rangle ^{(0)\,\,T}_{\rm R}  
\ena  
and $\mu,\nu\in\{0,...,5\}$ and $a\in\{6,...,9\}$.  
The zero modes part of the boundary state has a non trivial structure in both  
sectors; in the NS-NS case it is given by  
\beq  
\label{bound7}  
|B_{\psi} , \eta \rangle ^{(0)\,\,T}_{\rm NS}=  
\left(\hat C\frac{1+ i\eta\hat\gamma_5}{1+i\eta}  
\right)_{lm}|l\rangle|\tilde m\rangle  
\eeq  
where $\hat\gamma_i$ are the gamma matrices and $\hat C$ the charge   
conjugation matrix  
of $SO(4)$, $\hat\gamma_5=\hat\gamma_6...\hat\gamma_9$  
and finally  
$|l\rangle,|\tilde m\rangle$ are spinors of $SO(4)$, while  
in the R-R case we have  
\beq  
\label{bound8}  
|B_{\psi} , \eta \rangle ^{(0)\,\,T}_{\rm R}=  
\left(\bar C\bar\gamma^0...\bar\gamma^p\frac{1+ i\eta\bar\gamma_7}  
{1+i\eta}  
\right)_{ab}|a\rangle|\tilde b\rangle  
\eeq  
where analogously  
$\bar\gamma_i$ are the gamma matrices and $\bar C$ the charge   
conjugation matrix  
of $SO(1,5)$, $\bar\gamma_7=\bar\gamma_0...\bar\gamma_5$  
and finally  
$|a\rangle,|\tilde b\rangle$ are spinors of $SO(1,5)$.  
  
The GSO-projection selects  
both in the NS-NS and R-R twisted sectors  
the following combination \cite{gabstef}  
\beq  
\label{proi}  
|B\rangle^T=\frac{1}{2}\,\left[|B,+\rangle+|B,-\rangle\right]~~.
\eeq  
 The normalization constants  ${\cal N}^U$ and ${\cal N}^T$  
appearing in (\ref{bound1}) can be fixed by evaluating the  
interaction amplitude   
$\langle B|D|B\rangle$ between two fractional D$p$-branes  
due to the tree level exchange of closed strings  
and comparing the result with the   
corresponding open string channel calculation by means of   
the world-sheet duality transformation.  
The closed string amplitude can be written as the sum of two contributions,  
one for each sector  
\beq  
\label{int}  
\langle B|D|B\rangle=\,^U\langle B|D|B\rangle^U+\,^T\langle B|D|B\rangle^T  
\eeq 
An explicit computation easily shows  that   
the untwisted sector contribution   
vanishes due to the {\it abstruse identity},   
as in the case of  usual bulk brane, while  
in the twisted sector  the NS-NS and R-R contributions   
cancel each other.   
Therefore the amplitude  
eq.(\ref{int}) is vanishing and this is consistent with the fact     
that fractional branes, being BPS objects, satisfy a   
no-force condition.  
  
The  1-loop vacuum amplitude which represents the interaction between two  
fractional {\dpb} in the open  string channel  
is also expressed as the sum of two terms, one for  
each element of the orbifold group ${ Z_2}=\{1,g\}$.  
Under world-sheet duality they transform respectively   
in the two terms of eqs. (\ref{int}) and by comparison one can fix  
the normalization constants as follows
\beq  
\label{const}  
{\cal N}^U=\frac{T_p}{2\sqrt{2}}~~~~,~~~~  
{\cal N}^T=\frac{T_p}{2\sqrt{2}\pi^2\alpha'}  ~~.
\eeq  
  
\vskip 1cm  
\subsection{Brane couplings to bulk fields and their asymptotic behavior} 
\vskip 0.5cm  
  
In order to get information about the geometry of a fractional brane   
we need to determine its couplings with all the 
massless fields of the theory.  
This is achieved by projecting the boundary state   
onto the massless string states corresponding to the fields of the theory   
\cite{bs}.   
We need therefore a table of correspondence between the classical fields   
and the string states belonging to the massless spectrum.    
  
As already remarked, the action (\ref{ac10}) is a consistent truncation   
of the full action of type IIA supergravity in six dimensions, and   
describes the dynamics of the graviton $h_{\alpha\beta}$,   
four Kaluza-Klein scalars $\eta_a$ originating from the ten dimensional   
metric, a 1-form gauge field $C_1$, plus a 1-form gauge field $A_1$  
and a scalar $D$, which originates respectively from a three  
form gauge field and the  
Kalb-Ramond field with two components along the supersymmetric  
vanishing cycle of the orbifold.  
{F}rom the string analysis carried out in appendix {\bf B}, it is easy to   
realize that the graviton, the Kaluza-Klein scalars and the 1-form $C_1$  
are represented by the usual massless states of the untwisted sectors, while  
the 1-form $A_1$ and the scalar $D$ are described by massless states  
belonging, respectively, to the twisted R-R and NS-NS sector.   
In the case of the fractional D2-branes the twisted and untwisted 
1-forms are obviously replaced by two 3-forms ($C_3$ and $A_3$). 
 
By projecting the boundary state on the massless states corresponding 
to these fields   
(whose explicit expressions are given in appendix {\bf B},  
see also Ref.~\cite{eyras})  
we find the following couplings  
\beq  
\label{cogra}  
{\cal J}_h=  
-\frac{T_p}{\sqrt{2{\cal V}}} V_{p+1}\sum_{\a=0}^p h_\a^\a 
\eeq 
for the graviton, 
\beq  
\label{codil}  
{\cal J}_\phi=  
\frac{T_p}{2\sqrt{2{\cal V}}}V_{p+1}\phi(1-p)  
~~~~,~~~~
{\cal J}_{\eta_a}=  
\frac{T_p}{2 \sqrt {2{\cal V}}}V_{p+1}\eta_a  
\eeq  
for the dilaton and the Kaluza-Klein scalars,  
\beq  
\label{coru}  
{\cal J}_C= 
\frac{T_p}{\sqrt{\cal V}}V_{p+1}C_{0...p}  
\equiv\mu_U V_{p+1}C_{0...p}  
\eeq  
for the R-R untwisted field, and finally
\beq  
\label{cons}  
{\cal J}_{D}=  
-\frac{T_p}{2\pi^2\alpha'}V_{p+1}D  
~~~~,~~~~
{\cal J}_{A_{p+1}}=  
\frac{T_p}  
{2\pi^2\alpha'} V_{p+1}A_{0...p}\equiv\mu_T V_{p+1}A_{0...p}  
\eeq  
for the NS-NS twisted scalar and the R-R twisted field 
\footnote{Eqs. (\ref{cogra})-(\ref{cons}) 
can be extended to the case of the non-compact 
orbifold $C_2/\Gamma$  
by dropping the factor ${\cal V}$ in all equations.}.  
  
The previous couplings allow us to infer the structure of the   
Born-Infeld action for a fractional brane which turns out to be   
\[  
S_{wv}=-\frac{T_p}{\sqrt{2 {\cal V}}\kappa_{\rm orb}}  
\left\{\int d^{p+1}\xi   
\sqrt{-|g_{\alpha\beta}|}e^{-\kappa_{\rm orb}\phi\frac{(1-p)}{2}}  
\prod_a e^{-\kappa_{\rm orb}\frac{\eta_a}{2}}  
-\sqrt{2}\kappa_{\rm orb}\int C_{p+1}\right\}
\]  
\beq  
\label{bor}  
-\frac{T_p}{2\pi^2\alpha'}  
\left\{\int d^{p+1}\xi  
\sqrt{-|g_{\alpha\beta}|}e^{-\kappa_{\rm orb}\phi\frac{(1-p)}{2}}  
\prod_a e^{-\kappa_{\rm orb}\frac{\eta_a}{2}}D  
-\int \left[A_{p+1}+\sqrt{2}\kappa_{\rm orb}D \,C_{p+1}\right]  
\right\}   
\eeq  
The presence of the last term is due to the 
requirement of gauge invariance of  
the previous action under the gauge transformation 
\beq 
\label{gau} 
\delta A_{p+1}=\lambda_{p}\wedge dD~~~~,~~~~  
\delta C_{p+1}=\frac{1}{\sqrt{2}\kappa_{\rm orb}}d\lambda_{p} 
\eeq 
which  for $p=0$ is the gauge transformation  
that leaves invariant the bulk action (\ref{ac10}). 
The structure of the action (\ref{bor}) is confirmed also 
by explicit calculations of closed string scattering amplitudes 
on a disk with appropriate boundary conditions \cite{MERLATTI}. 
 
From the couplings in eqs.(\ref{cogra})-(\ref{cons}) one can also determine   
the contributions to the interaction between two fractional branes due   
to the exchange of each massless state.   
In particular, looking first at the untwisted states one finds  
the contribution of graviton  
\beq  
\label{intg}  
{\cal U}_h=  
\left(\frac{T_p}{2}\right)^2 \frac{V_{p+1}}{\cal V}
\frac{(p+1)(3-p)}{2 k^2_\perp}~~,
\eeq  
of the dilaton and Kaluza-Klein scalars  
\beq  
\label{ints}  
{\cal U}_\phi  
=\left(\frac{T_p}{2}\right)^2   
\frac{V_{p+1}}{\cal V}\frac{(1-p)^2}{2 k^2_\perp} ~~~~,~~~~
{\cal U}_{\eta_a}  
=\left(\frac{T_p}{2}\right)^2 \frac{V_{p+1}}{\cal V}\frac{1}{2 k^2_\perp}~~,
\eeq  
and of the R-R untwisted field  
\beq  
\label{intr}  
{\cal U}_C  
=-{T_p}^2 \frac{V_{p+1}}{\cal V}\frac{1}{k^2_\perp}~~.  
\eeq  
In the twisted sector instead, we find that the NS-NS scalar  
and the R-R gauge field  
contribute respectively as  
\beq  
\label{intns}  
{\cal U}_{D}  
=\left(\frac{{T_p}}{2\pi^2{\alpha'}}\right)^2 V_{p+1}\frac{1}{ k^2_\perp}  
~~~~,~~~~   
{\cal U}_{A_{(p+1)}}  
=-\left(\frac{{T_p}}{2\pi^2{\alpha'}}\right)^2 V_{p+1}\frac{1}{ k^2_\perp}~~.
\eeq  
The potential energies due to the exchange of the various fields   
satisfy the conditions  
\beq  
\label{nof1}  
{\cal U}_h+{\cal U}_\phi+\sum_a{\cal U}_{\eta_a}+{\cal U}_{C}=0  
~~~~,~~~~
{\cal U}_{A_{(p+1)}}+{\cal U}_{D}=0  
\eeq  
which are consistent with the fact that the no-force condition is due to  
the separate cancellation of the untwisted and twisted contributions to the  
interaction energy.  
  
The knowledge of the contribution to the interaction of each massless   
state is also useful to determine the mass of the fractional  
brane. Indeed in order to evaluate the mass we have to sum up all   
attractive contributions to the potential energy and compare it with   
Newton's law in six dimensions. By making a Fourier transformation 
to configuration space, we obtain that the attractive 
energy is   
\beq  
\label{massac}  
{\cal U}_{\rm attr.}= \frac{T_p^2}{(3-p)r^{3-p}\Omega_{4-p}}  
\left[\frac{1}{\cal V}+\frac{1}{(2\pi^2\alpha')^2}\right]  
\eeq  
where $\Omega_q=2\pi^{(q+1)/2}/\Gamma((q+1)/2)$ is the area of
a unit $q$-dimensional sphere.
Comparing eq.(\ref{massac}) with Newton's law  
\beq  
\label{newt}  
{\cal U}_{\rm Newt}=  
\frac{2\kappa^2_{\rm orb}{\cal M}_p^2}{(3-p)r^{3-p}\Omega_{4-p}}  
\eeq  
we get   
\beq  
\label{BPS}  
2\kappa^2_{\rm orb}{\cal M}_p^2={T_p}^2  
\left[\frac{1}{\cal V}+\frac{1}{(2\pi^2\alpha')^2}\right] 
=\left(\mu_U\right)^2+\left(\mu_T\right)^2~~, 
\eeq  
which is the usual relation between mass and charges of a BPS object   
charged with respect to two different gauge fields.  
  
Another important information provided by the boundary state   
analysis is the behavior of all classical fields generated   
by a fractional brane at large distance. This can be obtained   
by saturating the boundary state with the various states of the   
theory after inserting a closed string propagator.   
The asymptotic behavior for the various fields in our case is   
\beq  
\label{agra}  
h^\infty=\frac{T_p}{4\sqrt{2{\cal V}}}\frac{1}{(3-p)r^{3-p}\Omega_{4-p}}   
\left(\eta_{\alpha\beta}(p-3), \delta_{ij}(p+1)\right)  
\eeq  
for the graviton,
\beq  
\label{adil}  
\phi^\infty= 
\frac{T_p}{2\sqrt{2{\cal V}}} \frac{1-p}{(3-p)r^{3-p}\Omega_{4-p}}   
~~~~,~~~~  
{\eta_a}^\infty= 
\frac{T_p}{2\sqrt{2{\cal V}}} \frac{1}{(3-p)r^{3-p}\Omega_{4-p}}  
\eeq  
for the dilaton and the scalars,  
\beq  
\label{aru}  
C_{0...p}^\infty=-\frac{T_p}{\sqrt{\cal V}} \frac{1}{(3-p)r^{3-p} 
\Omega_{4-p}}   
\eeq  
for the untwisted R-R field, and finally  
\beq  
\label{ans}  
D^{\,\infty}=-\frac{T_p}{2\pi^2\alpha'}\frac{1}{(3-p)r^{3-p}\Omega_{4-p}}   
~~~~,~~~~  
{A_{0...p}^{\infty}}=-\frac{T_p}{2\pi^2\alpha'}\frac{1}{(3-p)r^{3-p}  
\Omega_{4-p}}   
\eeq  
for the twisted NS-NS scalar and the twisted R-R form.  
We remind that these asymptotic fields obtained with boundary state  
techniques have canonical normalization; therefore they do not   
coincide with the corresponding fields appearing in the bulk action  
given in eq.(\ref{ac10}).  
The relations between the two set of fields are   
\beq
\label{cano}  
\phi=\frac{\varphi}{\kappa_{\rm orb}}~~,~~  
\eta_a=\frac{\widetilde \eta_a}{\kappa_{\rm orb}}~~,~~  
C_{(p+1)}=\frac{{\widetilde C}_{(p+1)}}{\sqrt{2}\kappa_{\rm orb}}
~~,~~ 
D=\frac{{\tilde D}}{\kappa_{\rm orb}}~~,~~ 
{A_{(p+1)}}=\frac{{\widetilde A_{(p+1)}}}{\sqrt{2}\kappa_{\rm orb}}~~. 
\eeq
Using these new fields, the world-volume action of eq.(\ref{bor}) 
becomes 
\beq 
\label{act12} 
S_{wv}=-\frac{T_p}{\sqrt{2{\cal V}}\kappa_{\rm orb}}\left\{  
\int d^{p+1}\xi   
\sqrt{-|g_{\alpha\beta}|}\, e^{-\frac{\varphi(1-p)}{2}}  
\,\prod_a e^{-\frac{{\widetilde\eta}_a}{2}}  
-\int {\widetilde C}_{p+1}\right\}
\eeq 
\[  
-\frac{T_p}{2\sqrt{2}\pi^2\alpha'\kappa_{\rm orb}}\left\{  
\int d^{p+1}\xi \,\sqrt{2}\,  
\sqrt{-|g_{\alpha\beta}|}\, e^{-\frac{\varphi(1-p)}{2}}\prod_a  
e^{-\frac{{\widetilde\eta}_a}{2}}{\tilde D}  
-\int \left[{\widetilde A}_{p+1}+\sqrt{2}{\tilde D} 
{\widetilde C}_{p+1}\right]\right\}  ~.
\]  
This action will generate the source terms in the equations of
motions for the bulk fields.

\vskip 1.5cm  
\sect{Classical solution of fractional D-branes}  
\label{solution}  
\vskip 0.5cm  
  
In order to determine the classical solution of a fractional   
{\dpb} we can either solve the inhomogeneous field equations    
obtained by varying the total action {\it i.e.} the sum  of   
eqs.(\ref{ac10}) and (\ref{act12}),  
or solve the homogeneous field equations supplemented by   
the asymptotic behavior of fields, determined by the boundary state  
analysis, eqs. (\ref{agra})-(\ref{ans}). 
In appendix {\bf C} we find the classical solution associated to the  
fractional D$0$-brane following the latter procedure:  
we solve the homogeneous field equations under the assumption   
that all the fields are functions only of the radial transverse 
coordinate of the six-dimensional 
space and that the metric obeys the standard   
extremal black 0-brane {\it ansatz}.  
 
The first corrections to the asymptotic behavior of the fields can be easily  
obtained by solving iteratively the equations of motion.   
From this analysis  one can learn that, up to second order,  
{\it the twisted fields   
do not get corrections with respect to their harmonic asymptotic behavior.}   
This appears to be  an important physical feature of the twisted fields.   
Therefore  we  
take it as an {\it ansatz} for the full solution.  
Once made this assumption the equations of motion are   
easily solved in terms of a single function $H$, which is a very simple  
generalization of the harmonic function appearing in the classical solution  
of ordinary bulk branes  
\beq  
\label{s3}  
H=1+\frac{1}{2}\frac{Q_0}{r^{3}}
-\frac{1}{2}\frac{{\cal V}}{(2\pi)^4\alpha'^2}  
\frac{Q_0^2}{r^{6}} ~~~~, 
\eeq  
where 
\beq  
\label{Q0}  
Q_0=\frac{2\sqrt{2}T_0\kappa_{\rm orb}}{3\Omega_4{\cal V}^{1/2}} ~~~~. 
\eeq  
More explicitly we have
\beq  
\label{s4}  
{ds}^2 = - H^{-\frac{3}{4}} dt^2  + H^{\frac{1}{4}}(dr^2+r^2d\Omega^2)  
\eeq  
for the metric,  
\beq  
\label{s5}  
e^\varphi=H^{\frac{1}{4}}  
~~~~,~~~~   
e^{\eta_a}=H^{\frac{1}{4}}
\eeq  
for the dilaton and for the scalar fields,  
\beq  
\label{s7}  
C_{0}= \left( H^{-1}-1 \right)  
\eeq  
for the untwisted R-R vector, and finally 
\beq  
\label{s8}  
D=-\frac{1}{\sqrt 2}\frac{{\cal V}^{1/2}}{4\pi^2\alpha'}\frac{Q_0}{r^{3}}  
~~~~,~~~~   
A_{0} =-\frac{{\cal V}^{1/2}}{4\pi^2\alpha'}\frac{Q_0}{r^{3}}  
\eeq  
for the twisted NS-NS scalar and the twisted R-R vector, respectively.  
 
The structure of the D$0$-brane solution is very simple: for the twisted  
fields the first order correction to the background value is exact,  
while the untwisted fields have the same expression of the fields  
associated to a bulk brane in terms of the function $H$. 
This fact suggests a natural generalization of our solution to the  
case of the fractional D$2$-brane, namely 
\beq  
\label{s4b}  
{ds}^2 =H^{-\frac{1}{4}} (-dt^2+dx^2_1+dx^2_2)  + 
H^{\frac{3}{4}}(dr^2+r^2d\Omega^2)  
\eeq  
for the metric,  
\beq  
\label{s5b}  
e^\varphi=H^{-\frac{1}{4}}  
~~~~,~~~~
e^{\eta_a}=H^{\frac{1}{4}}
\eeq  
for the dilaton and for the scalar fields, 
\beq  
\label{s7b}  
C_{012}= \left( H^{-1}-1 \right)  
\eeq  
for the untwisted R-R vector and   
\beq  
\label{s8b}  
D=-\frac{1}{\sqrt 2}\frac{{\cal V}^{1/2}}{4\pi^2\alpha'}\frac{Q_2}{r}  
~~~~,~~~~
A_{012} =-\frac{{\cal V}^{1/2}}{4\pi^2\alpha'}\frac{Q_2}{r}  
\eeq  
for the twisted NS-NS scalar and for the twisted R-R vector  
respectively, where   
\beq  
\label{sh}  
H=1+\frac{1}{2}\frac{Q_2}{r}-\frac{1}{2} 
\frac{{\cal V}}{(2\pi)^4\alpha'^2}\frac{Q_2^2}{r^{2}}  
\eeq  
and  
\beq  
\label{Q2}  
Q_2=\frac{2\sqrt{2}T_2\kappa_{\rm orb}}{\Omega_2{\cal V}^{1/2}} ~~~. 
\eeq  
 
Let us make some comments about the solution both for $p=0$ and $p=2$. 
First of all one can notice that this solution is consistent with the BPS   
nature of fractional {\dpbs}, as seen from the mass-charge relation  
(\ref{BPS}) and from the one loop no-force condition, verified by the  
boundary state technique. 
In fact, by computing the world-volume action (\ref{act12}) for a probe  
fractional brane in the background of Eqs. (\ref{s4}-\ref{s8}) or  
Eqs.(\ref{s4b}-\ref{s8b}), one can check that the distance dependent part  
identically vanishes, and therefore there is no static force acting on  
the probe.

Another important observation is that the untwisted fields, and the metric  
in particular, have a naked singularity at $r=r_{+}$ where  
\beq 
\label{} 
(r_{+})^{3-p} =\frac{Q_p} {4}\left(-1 +\sqrt{1+ 
\frac{{\cal V}}{2\pi^4\alpha '^2}}\,\right) ~~~~,~~~p=0,2 ~~. 
\label{r+} 
\eeq 
However the metric we have found is strictly analogous to the one studied   
in Ref.~\cite{JPP}, so we expect that this singularity is of {\it repulson}  
type and one may cure it by means of an {\it enhan\c{c}on} mechanism first  
proposed therein.  
To see this, one notices that at a distance $r_e$, where    
\beq 
(r_e)^{3-p} = 2  {Q}_p \frac{{\cal V}}{{(2 \pi)^4\alpha '}^2} 
>(r_+)^{3-p} 
\label{re} 
\eeq 
the derivative of $H$ vanishes and both gravitational and gauge forces 
change sign. 
Therefore, even if a fractional {\dpb} probe feels no net force at any   
distance, it  becomes tensionless at $r=r_e$ and acquires a negative  
tension at shorter distances. Indeed expanding the DBI part of the  
world-volume action in the velocities and keeping only the lowest order  
terms one gets   
\begin{eqnarray}
-\,\frac{T_p}{\sqrt{2{\cal V}}\kappa_{\rm orb}} 
\int d{\xi}^{p+1}{\delta}_{ij}  
\frac{\partial x^i}{\partial \tau }  
\frac{ \partial x^j}{\partial \tau}   
\left(1+\frac{\sqrt{2{\cal V}}}{2\pi^2\alpha'}D\right )= ~~~~~~~~~~~~~~~~~~~
\nonumber \\
\label{vel2}  
~~~~~~~~~~~~~~~~~~~~~~~~
-\,\frac{T_p}{\sqrt{2{\cal V}}\kappa_{\rm orb}}\int d{\xi}^{p+1}{\delta}_{ij}  
\frac{\partial x^i}{\partial \tau }  
\frac{ \partial x^j}{\partial \tau}   
\left(1- \frac{r^{3-p}_e}{r^{3-p}}\right ) ~~.
\end{eqnarray}
We see then that if $D$ has the form of eq.(\ref{s8}) or (\ref{s8b}) 
the {\dpb}  becomes tensionless exactly at $r=r_e$.  
It is then clear that the classical solution   
cannot be trusted at $r <r_e$.  
This may correspond to the fact that, as shown in Ref. \cite{JPP}, 
the {\dpbs} building up the classical background,
rather than piling up at $r=0$, are forced to cover   
uniformly the hypersphere at $r=r_e$.   
If one identifies the (properly rescaled) transverse coordinates with 
the Higgs fields, one may use the procedure just outlined to derive their 
kinetic term.  
Then, for the $(2+1)$ gauge theory dual to the D$2$ fractional brane,  
the factor ${\delta}_{ij} ( 1- r_e/r)$ can be interpreted as  
the metric in moduli space, which turns out to be rather independent  
from the detailed geometry of D$2$-brane.  
Actually from this point of view the fact that NS-NS twisted   
field keeps its harmonic asymptotic form   
may be seen as expressing in geometrical classical terms  
the quantum property of ${\cal N}=2$ SUSY theory allowing only one-loop    
perturbative corrections. It is then worth investigating more closely the  
correspondence between the detailed structure of fractional {\dpb} geometry   
and the associated world-volume gauge theory.

\vskip 3.5cm  
{\large {\bf Acknowledgments}}  
\vskip 0.5cm  
\noindent  
We would like to thank M. Bill\'o, M. Bertolini,  
P. Di Vecchia,  L. Gallot, A. Lerda, R. Marotta, I. Pesando, F. Pezzella   
and R. Russo for very useful discussions. A.L. acknowledges  
support from Fondazione Angelo Della Riccia, and thanks NORDITA  
and Universit\`a degli Studi di Napoli   
for kind hospitality.  
  
\appendix
\vskip 1.5cm  
\section{Alternative derivation of the bulk action}  
\label{appea}  
\vskip 0.5cm  
\renewcommand{\theequation}{A.\arabic{equation}}  
\setcounter{equation}{0}  
\noindent  
In this appendix we give an alternative derivation of the   
truncated bulk action necessary to reconstruct the geometry of a fractional 
brane.  
This alternative strategy consists in exploiting the   
$S$-duality relating type IIA on $T_4/I_4$  
to  heterotic theory compactified on $T_4$.   
Furthermore our task is made easier by turning on  only the fields which  
couple to fractional {\dpbs} at  tree level. For the sake of definiteness  
we refer here to the case of {\db} where the fields are  
the graviton $g_{\mu\nu}$, 4 scalar fields $\eta_a$,  
the dilaton $\phi$, 1 untwisted R-R vector $C_0$,  
16 twisted R-R vectors $A_0^I$ ($I=1,\ldots,16$)
and 16 twisted NS-NS scalars $D^I$.  
 
One may easily check that under the chain of dualities   
(implementing  the 6 dimensional $S$-duality  from   
IIA on $T_4/I_4$ to the heterotic string on $T_4$)  
 \[  
IIA \,\,\,\frac{T_4}{I_4} \stackrel{T_9}{\Longleftrightarrow} IIB \,\,\,  
\frac{T_4}{(-1)^{F_L}I_4}   
\stackrel{S}{\Longleftrightarrow}IIB\,\,\, \frac{T_4}{\Omega I_4}   
\]  
\beq  
\label{catena}  
\stackrel{4\,\,T}{\Longleftrightarrow}  
{\rm Type \,I}\,\,\, T_4 \stackrel{S}{\Longleftrightarrow} 
{\rm heterotic}\,\,\,T_4  ~~,
\eeq  
the previous fields transform as follows  
\beq  
\label{tra1}  
{\varphi}={-\varphi^{\rm he}}~~~~,~~~~  
g_{\mu\nu}=e^{-2\varphi_{\rm he}} g_{\mu\nu}^{\rm he}~~,  
\eeq  
\beq  
\label{tra2}  
G_{\,aa}=\left(\prod_{b}G_{\,bb}^{\rm he}\right)^{1/2}
\left(G_{\,aa}^{\rm he}\right)^{-1}   
~~~{\rm for}~~ a\neq 9~~,~~{\rm and}~~  
G_{\,99}=\left(\prod_{b}G_{\,bb}^{\rm he}\right)^{-1/2}G_{\,99}^{\rm he}~~,  
\eeq  
\beq  
\label{tra3}  
C_{\,\mu}= G_{\,\mu 9}^{\rm he} 
G_{\rm he}^{\,99}\equiv A_{\,\mu}^{9\,{\rm he}}~~~~,~~~~  
(A_{\,\mu}^I)= 
\frac{A_{\,\mu}^{I\,{\rm he}}+A_{\,\mu}^{I+1\,{\rm he}}}{\sqrt{2}}   
\eeq  
\beq  
\label{tra4}  
D^I=\frac{A_9^{I\,{\rm he}}+A_9^{I+1\,{\rm he}}}{\sqrt{2}}   
\eeq  
where the label ``he'' refers to heterotic fields. 

The effective action of the 
heterotic string compactified on a 4-torus of volume
$\mu^4$ is given by  
\[  
S^{\rm he} = \frac{\mu^4}{2 \kappa_{10}^2}  
\int d^6 x\sqrt{-g^{\rm he}}\,e^{-2\varphi^{\rm he}}  \,
\left[{R}(g^{\rm he}) + 4\partial_\mu\varphi^{\rm he}
\partial^\mu\varphi^{\rm he} +   
\frac{1}{4}\left(\partial_\mu {G_{ab}}\partial^\mu {G^{ab}}\right)^{\rm he}  
\right.  
\]  
\[  
\left.  
-\frac{1}{4}\left(G_{ab}F_{\mu\nu}^{\,\,a} F^{\mu\nu b}\right)^{\rm he}  
-\frac{1}{12}\left(h_{\mu\nu\rho}h^{\mu\nu\rho}\right)^{\rm he} 
-\frac{1}{4}\left(G_{ab}h_{\mu\nu}^{\,\,a} h^{\mu\nu b}\right)^{\rm he}  
-\frac{1}{4}\left(G_{ab}G_{cd}h_{\mu}^{\, ac} h^{\mu bd}\right)^{\rm he} 
\right]  
\]  
\beq  
-\frac{\mu^4}{4 g_{10}^2}  
\int d^6 x \sqrt{-g^{\rm he}}\,e^{-2\varphi^{\rm he}}\,  
\sum_{I=1}^{16}\left(\widetilde F_{\mu\nu}^I\widetilde F^{I\mu\nu} 
+2G_{ab}\widetilde F_{\mu}^{\,a\, I} \widetilde F^{\mu b\,I}\right)^{\rm he}  
\label{eter1}  
\eeq  
where we have assumed that the heterotic gauge group is broken 
to $U(1)^{16}$ and have defined the field  
$\widetilde F_{\mu\nu}^{I\,\rm he}$ follows  
\beq  
\widetilde F_{\mu\nu}^{I\,{\rm he}}=
\left(F_{\mu\nu}^{I}+A^a_{[\mu}\partial_{\nu]} A^{I}_{\,a}\right)^{\rm he} ~~. 
\eeq
To make contact with the case considered in the previous sections,
we first neglect in the action (\ref{eter1}) all terms that
contain fields not dual to those 
of the truncated type IIA theory we are interested in, which
contains only one pair
of twisted fields (say $(A^J,D^J)$ for a given $J$).
Doing this, we get  
\[  
S^{\rm he} = \frac{\mu^4}{2 \kappa_{10}^2}  
\int d^6 x\sqrt{-g^{\rm he}}\,e^{-2\varphi^{\rm he}}\,  
\Bigg[R(g^{\rm he}) + 
4\partial_\mu\varphi^{\rm he}\partial^\mu\varphi^{\rm he} 
\]  
\[  
+\frac{1}{4}\left(\partial_\mu {G_{aa}}\partial^\mu {G^{aa}}\right)^{\rm he}  
-\frac{1}{4}\left(G_{99}F_{\mu\nu}^9 F^{\mu\nu 9}\right)^{\rm he}  
\Bigg]  
\]  
\beq  
-\frac{\mu^4}{4 g_{10}^2}  
\int d^6 x \sqrt{-g^{\rm he}}\,e^{-2\varphi^{\rm he}}  
\sum_{I=J, J+1}\left(\widetilde F_{\mu\nu}^I \widetilde F^{I\mu\nu}  
+2G_{99}\widetilde F_{\mu}^{I\,9}\widetilde F^{\mu 9\,I}\right)^{\rm he}  ~~.
\label{eter2}  
\eeq  
By performing the $S$-duality transformations (eqs.(\ref{tra1})-(\ref{tra4}))  
on the previous action,
we get the following truncated low energy  action for IIA on $T_4/I_4$:  
\[  
S= \frac{\mu^4}{2 \kappa_{10}^2}  
\int d^6 x\sqrt{-g}\,e^{-2\varphi} \, 
\Bigg[R(g) + 4\partial_\mu\varphi\partial^\mu\varphi -   
\partial_\mu {\widetilde\eta_{a}}\partial^\mu {\widetilde\eta^{a}}   
\]  
\[  
\left.  
-\frac{1}{4}\prod_a e^{\widetilde\eta_a}F_{\mu\nu}F^{\mu\nu}  
\right]  
-\frac{\mu^4}{4 g_{10}^2}  
\int d^6 x \sqrt{-g}\,  
\Bigg[F_{\mu\nu}^J F^{J\mu\nu}   
\]  
\beq  
\left.  
+ 2F_{\mu\nu}^J C^{[\mu}\partial^{\nu]}D^J+  
C_{[\mu}\partial_{\nu]}D^J C^{[\mu}\partial^{\nu]}D^J  
+2 e^{-2\varphi}\partial_\mu D^J\partial^\mu D^J 
\prod_{a}e^{-\widetilde\eta^a}  
\right] 
\label{2a}  
\eeq  
where $F_{\mu\nu}=\partial_{[\mu}C_{\nu]}$ and
$F_{\mu\nu}^J=\partial_{[\mu}A_{\nu]}^J$. 
After extracting the dilaton vacuum expectation value,   
one may rewrite the previous action in the Einstein frame as follows:  
\[  
S= \frac{1}{2 \kappa_{\rm orb}^2}  
\int d^6 x\sqrt{-g}  
\left[{R}(g) -\partial_\mu\varphi\partial^\mu\varphi   
-\partial_\mu {\widetilde\eta_{a}}\partial^\mu {\widetilde\eta^{a}}  
-\frac{1}{4}e^{\varphi}  
\prod_a e^{\widetilde\eta_a}{\widetilde F}_{\mu\nu}{\widetilde F}^{\mu\nu}  
\right.  
\]  
\[  
\left.  
-\frac{1}{4}  
e^\varphi  \,
{\widetilde F}_{\mu\nu}^J 
{\widetilde F}^{J\mu\nu} -
\frac{1}{\sqrt 2}e^{\varphi}{\widetilde F}_{\mu\nu}^J 
{\widetilde C}^{[\mu}\partial^{\nu]}{\tilde D}^J-
\frac{1}{2}e^{\varphi}  
{\widetilde C}_{[\mu}\partial_{\nu]}{\tilde D}^J 
{\widetilde C}^{[\mu}\partial^{\nu]}{\tilde D}^J  
\right.  
\]  
\beq  
\left.  
-\partial_\mu {\tilde D}^J\partial^\mu {\tilde D}^J \prod_{a} 
e^{-\widetilde\eta_a}  
\right]  
\label{2aa}  
\eeq  
where $\kappa_{\rm orb}$ has been defined in section {\bf 2} and     
$g^2_{\rm orb}= {g^2_{10}}/{\mu^4}$.  
Moreover the following rescalings on the fields   
in eq.({\ref{2a}}) have been made:  
\beq  
{\widetilde C}_{\mu}=\frac{g_s\mu^2}{\prod_a(2\pi R_a)^{1/2}}C_\mu
~~~~,~~~~  
{\widetilde A}_{\mu}^J
=\frac{\sqrt{2}\kappa_{orb}}{g_{\rm orb}}A_\mu^J~~~~,~~~~  
{\tilde D}^J=\frac{\kappa_{\rm orb}
\prod_a(2\pi R_a)^{1/2}}{g_{\rm orb}g_s\mu^2}D^J~~.  
\label {ridef}  
\eeq  
Under the identifications 
\[ 
\tilde D=-\tilde D^J ~~~~,~~~~  
\tilde A_\mu=\tilde A_\mu^J 
\]  
the action (\ref{2aa}) coincides with the one obtained in section {\bf 2} 
with a different procedure.  

\vskip 1.5cm  
\section{Untwisted and twisted states}  
\label{appeb}  
\vskip 0.5cm  
\renewcommand{\theequation}{B.\arabic{equation}}  
\setcounter{equation}{0}  
\noindent  
  
In this appendix we  briefly review the  description of the orbifold theory   
we are interested   
in, namely   type IIA compactified on $T_4/{Z_2}$,  and  its spectrum.   
 
In a generic orbifold theory   
the periodicity conditions of the closed string  
are satisfied up to the action of a generic element of the orbifold group.  
Thus in the case  of  the discrete group ${Z}_N$   
which consists of $N$  elements $\{1, g,...,g^{N-1}\}$  
we may have  
\beq  
\label{twib}  
X(\tau,\sigma+\pi)=g^m X(\tau,\sigma)g^{-m}\,\,\,\,\,
{\rm with}\,\,\,\,\,\,m\in\{0,...,N-1\}  
\eeq  
For $m=0$ one recovers the  boundary 
conditions of closed string without orbifold projection,  
called untwisted boundary conditions, while for $m\neq 0$ one has $N-1$  
different twisted boundary conditions.  
In the case of a ${Z}_2$ projection,  acting  as a   
reflection over four space coordinates,  
the ten-dimensional  
Lorentz group   
$SO(1,9)$ decomposes in $SO(1,5)\times SO(4)$.  
We fix our convention as follows:  
$M,P,Q$ are the ten-dimensional indices;    
$a,b\in\{6,7,8,9\}$  are the indices corresponding 
to the coordinates which are reflected by $I_4$ and  
$\mu,\nu\in\{0,...,5\}$  are the directions 
transverse to the orbifold projection.   
The theory has a unique    
twisted sector for each one of the sixteen orbifold fixed planes,   
characterized by the following boundary conditions   
\beq  
\label{twibb}  
X^\mu(\tau,\sigma+\pi)=X^\mu(\tau,\sigma)\,\,\,\,\,{\rm and}
 \,\,\,\,X^a(\tau,\sigma+\pi)=-X^a(\tau,\sigma)  
\eeq  
for the bosonic coordinates and  
\beq  
\label {ucbc}  
\psi^\mu_\pm(0,\tau)=\psi^\mu_\pm(\pi,\tau)~~;~~  
\psi^a_\pm(0,\tau)=-\psi^a_\pm(\pi,\tau)  
\eeq  
for the R-R twisted sector and  
\beq  
\label {ucbc'}  
\psi^\mu_\pm(0,\tau)=-\psi^\mu_\pm(\pi,\tau)~~;~~  
\psi^a_\pm(0,\tau)=\psi^a_\pm(\pi,\tau)  
\eeq  
for the NS-NS twisted sector.  
Obviously different boundary conditions correspond to different 
mode expansions   
of the string coordinates.  
The bosonic string mode expansion of the twisted sector is given by  
\beq  
\label {expc}  
X^\mu(\tau, \sigma)= q^\mu+2\alpha'p^\mu \tau +i\sqrt{\frac{\alpha'}{2} }  
\sum_{n\neq 0}  
\left(\frac {\alpha^{\mu}_{n}}{n}e^{-2in(\tau -\sigma)}+  
\frac {\tilde\alpha^{\mu}_{n}}{n}e^{-2in(\tau+\sigma)}\right) 
\eeq  
for each $\mu\in\{0,...,5\}$ and  
\beq  
\label {expc1}  
X^a(\tau, \sigma)= i\sqrt{\frac{\alpha'}{2} }  
\sum_{r=1/2}^\infty  
\left(\frac {\alpha^{a}_{r}}{r}e^{-2ir(\tau -\sigma)}+  
\frac {\tilde\alpha^{a}_{r}}{r}e^{-2ir(\tau+\sigma)}\right)
\eeq  
for each $a\in \{6,,,.9\}$.  
For the fermionic coordinates one finds  
\beq  
\label{modpsil2}  
\psi^M_-=\sqrt{\alpha'}\sum_{t} \psi^M_t  
e^{-2it(\tau-\sigma)}~~~~~~~\psi^M_+=\sqrt{\alpha'}\sum_{t} \tilde\psi^M_t   
e^{-2it(\tau +\sigma)}  
\eeq  
where  
\beq  
\label{modpsil3}  
\left\{   
\begin{array}{l}  
\psi^\mu_t \,\,\, {\rm and}\,\,\,\tilde\psi^\mu_t\,\,\,\, t\in Z\\  
\psi^a_t\,\,\,\,{\rm and}\,\,\,\tilde\psi^a_t\,\,\,\, t\in Z+\frac{1}{2}\\  
\end{array}  
\right. ,  
\rightarrow {\rm R-R \,\,\, twisted \,\,\,sector}  
\eeq  
and  
\beq  
\label{modpsil4}  
\left\{   
\begin{array}{l}  
\psi^\mu_t \,\,\,\,{\rm and}\,\,\,\tilde\psi^\mu_t\,\,\,\,
t\in Z+\frac{1}{2}\\  
\psi^a_t\,\,\,\,{\rm and}\,\,\,\tilde\psi^a_t\,\,\,\, t\in Z\\  
\end{array}  
\right. ,  
\rightarrow {\rm NS-NS \,\,\, twisted \,\,\,sector}  
\eeq  
To construct the massless spectrum of the theory one has to impose the   
mass-shell   
condition separately in each sector. For the sake of simplicity   
we considere a  4-torus of the type   
$T_4=T_1\times T_1\times  T_1\times T_1$; in this case
the mass-shell condition in the twisted NS-NS sector reads as  
follows   
\beq  
\label{masscon}  
\left\{  
\frac{2}{\alpha'}\left(\tilde N_\alpha+\tilde 
N_\psi+ N_\alpha+N_\psi-1\right)+\sum_{a}\left[  
\left(\frac{n_a}{R_a}\right)^2 +\left(\frac{w_a R_a}{\alpha'}\right)^2\right]  
\right\}|{\rm state}\rangle=0  
\eeq  
and must be imposed together with the level matching condition  
\beq  
\label{lema}  
(\tilde N_\alpha+\tilde N_\psi- N_\alpha-N_\psi)|{\rm state}\rangle= 
\sum_a n_a w_a|{\rm state}\rangle  ~~.
\eeq  
with $N_\alpha$ and $N_\psi$ being the bosonic and fermionic occupation  
numbers, $ R_a$  the compactification radius of the coordinate $X_a$  
and finally $n_a$ and $w_a$  respectively   
Kaluza-Klein and  winding modes along the compact direction $a$.  
For $R\neq \sqrt{\alpha'}$ it turns out 
that the massless states in the NS-NS untwisted sector are  
the graviton which 
transforms as $(3,3)$ under the action of the little group $SO(4)$,
the Kalb-Ramond field (($1,3)$ + $(3,1)$), the dilaton ($(1,1)$),  
10 Kaluza-Klein scalars coming from the compactification of the 
ten-dimensional  
graviton ($10(1,1)$), and 6 scalars coming from  
the compactification of the Kalb-Ramond field ($6(1,1)$).  
The explicit expressions of these states are  
\begin{eqnarray}  
\ket{\Psi_h} &=& h_{\mu \nu}~{\tilde{\psi}}^{ \mu}_{-\frac{1}{2}}   
\psi^{\nu}_{-\frac{1}{2}}~  
~\ket{0_\psi}_{-1}\, \tilde{\ket{0_{\tilde\psi}}}_{-1}\ket{k}
\prod_{a}\frac{1}{\sqrt{{\bf\Phi}_a}}\ket{n_a=w_a=0}~~,  
\label{etai}\\  
\ket{\Psi_B} &= &\frac{B_{\mu \nu}}{\sqrt{2}}
~{\tilde{\psi}}^{ \mu}_{-\frac{1}{2}}   
\psi^{ \nu}_{-\frac{1}{2}}~\ket{0_\psi}_{-1}\, 
\tilde{\ket{0_{\tilde\psi}}}_{-1}\ket{k}\prod_{a}\frac{1}{\sqrt{{\bf\Phi}_a}}
\ket{n_a=w_a=0}~~,  
\label{anti}\\  
\ket{\Psi_\phi} &=&\frac{\phi}{\sqrt{8}} \left(\eta_{\mu\nu} -   
k_{\mu} \ell_{\nu} -  k_{\nu}\ell_{\mu}\right)\,  
{\tilde{\psi}}^{\mu}_{-\frac{1}{2}} \psi^{\nu}_{-\frac{1}{2}}   
~\ket{0_\psi}_{-1}\, \tilde{\ket{0_{\tilde\psi}}}_{-1}\ket{k} 
\nonumber 
\\ 
&&\prod_{a}\frac{1}{\sqrt{{\bf\Phi}_a}}\ket{n_a=w_a=0}~~,  
\label{dila} \\  
\ket{\Psi_{\eta_a}} &=& \eta_a~{\tilde{\psi}}^{ a}_{-\frac{1}{2}}   
\psi^{a}_{-\frac{1}{2}}~  
~\ket{0_\psi}_{-1}\, \tilde{\ket{0_{\tilde\psi}}}_{-1}\ket{k}
\prod_{a}\frac{1}{\sqrt{{\bf\Phi}_a}}\ket{n_a=w_a=0}~~,  
\label{grav}\\  
\ket{\Psi_{B_{ab}}} &=&
\frac{B_{ab}}{\sqrt{2}}~{\tilde{\psi}}^{a}_{-\frac{1}{2}}   
\psi^{b}_{-\frac{1}{2}}~\ket{0_\psi}_{-1}\, 
\tilde{\ket{0_{\tilde\psi}}}_{-1}\ket{k}\prod_{a}\frac{1}{\sqrt{{\bf\Phi}_a}}
\ket{n_a=w_a=0}~~,  
\label{antis}  
\end{eqnarray}  
where ${\bf\Phi}_a$ is the self-dual volume of the compact direction 
$x_a$ \cite{give,noi}.  
Moreover $h_{\mu\nu}$, $B_{\mu\nu}$ and $\phi$  
are  graviton, Kalb-Ramond and dilaton  field  
respectively, while $B_{ab}$ and $\eta_a$ are  the scalars   
corresponding to the compact components of  Kalb-Ramond field  
and to the compact diagonal components  
of the metric.   
In the R-R untwisted sector instead the mass-shell condition reads as 
follows   
\beq  
\label{massconr}  
\left\{  
\frac{2}{\alpha'}\left(\tilde N_\alpha+\tilde N_\psi+ N_\alpha+
N_\psi\right)+\sum_{a}\left[  
\left(\frac{n_a}{R_a}\right)^2 +\left(\frac{w_a R_a}{\alpha'}  
\right)^2\right]\right\} |{\rm state}\rangle=0  
\eeq  
and it should be imposed together with the level-matching 
 condition given in eq.(\ref{lema}).  
In type IIA theory compactified   
on $T_4/{Z_2}$ the massless states turn out to be 8 vectors 
transforming as $(2,2)$  
under the action of the little group $SO(4)$, whereas in 
type IIB theory on $T_4/Z_2$ one find 8 scalars, and    
3 2-forms potential, transforming respectively as 
$(1,1)$, and $((3,1)+(1,3))$ under the little group.  
Their explicit expression is given by  
\begin{eqnarray}  
\label{fred6}  
\ket{C_{(n)}} & = & \frac{1}{2\sqrt{2}\,n!}~  
C_{\mu_1\ldots\mu_n}\Bigg[  
\left(C\Gamma^{\mu_1\ldots\mu_n}\Pi_+\right)_{AB}\,  
\cos (\gamma_0{\tilde\beta}_0)  
\\  
&&~+~  
\left( C\Gamma^{\mu_1\ldots\mu_n}\Pi_-\right)_{AB}\,  
\sin (\gamma_0{\tilde\beta}_0) \Bigg]  
\ket{A}_{-1/2}~\ket{\tilde B}_{-3/2}\ket{k}  
\nonumber  
\end{eqnarray}  
where $\Pi_\pm = (1\pm\Gamma_{11})/2$, and $\beta_0$ and $\gamma_0$  
are the superghost zero-modes.  
  
Let us examine now explicitly the twisted mass spectrum.  
In this case the mass-shell condition is independent of  
the Kaluza-Klein and winding modes and it reduces to  
\beq  
\label{masst}  
\frac{2}{\alpha'}\left(\tilde N_\alpha+\tilde N_\psi+ 
N_\alpha+N_\psi\right)|{\rm state}\rangle=0  
\eeq  
both in the NS-NS and R-R twisted sectors.  
The NS-NS twisted states are a product of two spinors  
of the internal $SO(4)$ having the same chirality  
(this is true both in type IIA and type IIB theory on $T_4/{Z_2}$).   
Therefore, from the point of view of the  
internal space, these states transform  
as $(2,1)\otimes(2,1)=(3,1)+(1,1)$ which corresponds  
to the transformation properties of a self dual 2-form  
that we denote by  $D_{ab}^I$ and a scalar  that we denote by  
$D^I$, where the index $I$   
runs from $1$ to $16$.   
All these states transform as scalars under the little group,  
thus there are  $4\times 16$ scalars more in the theory.  
  
Following the same notations of sect. 3 the explicit expression of these  
states is  
\begin{eqnarray}  
\label{fred61}  
\ket{D_{(n)}^I} & = & \frac{1}{\sqrt{2}\,n!}~  
D_{a_1\ldots\a_n}^I 
\left(\hat C\hat\gamma^{\a_1\ldots\a_n}\hat\Pi_+\right)_{lm}\,  
\ket{l}_{-1}~\ket{\tilde m}_{-1}\ket{k}  
\nonumber  
\end{eqnarray}    
Finally, in the R-R twisted sectors,   
one finds only one massless state for each fixed point which is a vector  
in type IIA and an antiself-dual 2 form plus a scalar in type IIB theory.   
These states have the following expression  
\begin{eqnarray}  
\label{fred7}  
\ket{A_{\mu_0,...,\mu_m}^I} & = & \frac{1}{\sqrt 2}~  
\frac{A_{\mu_0,...,\mu_m}^I}{m!}\Bigg[  
\left(\bar C\bar\gamma^{\mu_0,...,\mu_m}\bar\Pi_+\right)_{ab}\,  
\cos(\gamma_0{\tilde\beta}_0)  
\\  
&&~+~  
\left( \bar C\bar\gamma^{\mu_0,...,\mu_m}\bar\Pi_-\right)_{ab}\,  
\sin (\gamma_0{\tilde\beta}_0) \Bigg]  
\ket{a}_{-1/2}~\ket{\tilde b}_{-3/2}\ket{k}  
\nonumber  
\end{eqnarray}  
  
\vskip 1.5cm  
\section{Field equations}  
\label{appec}  
\vskip 0.5cm  
\renewcommand{\theequation}{C.\arabic{equation}}  
\setcounter{equation}{0}  
\noindent  
In this appendix we write down  and  
solve the homogeneous equations of motion and fix all the    
integration constants in order to obtain a solution consistent with  
the large distance behavior of all the fields.   
For the sake of simplicity  
  let us start by rewriting in components    
the low-energy action obtained in section {\bf 2}
\[  
S= \frac{1}{2 \kappa_{\rm orb}^2}  
\int d^6 x\sqrt{-g}  
\left\{{R}(g) -\partial_\mu\varphi\partial^\mu\varphi   
-\sum_{a=6}^9\partial_\mu {\eta_{a}}\partial^\mu {\eta^{a}}  
-\frac{e^{\varphi}}{4}\left[  
\prod_a e^{\eta_a}{F}_{\mu_1\mu_2}{F}^{\mu_1\mu_{2}}  
\right.  
\right.  
\]  
\[  
\left.  
+  
{F}_{\mu_1\mu_2}^A {F}_A^{\mu_1\mu_{2}} -  
2\sqrt{2}{F}_{\mu_1\mu_2}^A   
{C}^{[\mu_1}\partial^{\mu_{2}]}  
{D}  
\right.  
\]  
\beq  
\left. 
\left. 
+2  
{C}_{[\mu_1}\partial_{\mu_{2}]}  
D {C}^{[\mu_1}\partial^{\mu_{2}]}{D}\right]  
-\partial_\mu {D}\partial^\mu {D} \prod_{a}e^{-\eta_a}  
\right\}  
\label{2ab}  
\eeq  
where we have
dropped  all tildas from the fields, to simplify the notation.  
The corresponding equations of motion read as follows:    
\[  
\frac{2}{\sqrt{-g}}\partial_{\mu}  
\left(\sqrt{-g}\partial^\mu \varphi\right)=  
\frac{1}{4}e^{\varphi}\left[\prod_a e^{\eta_a}  
F_{\mu_1\mu_{2}}F^{\mu_1\mu_{2}}+  
\right.  
\]  
\beq  
\label{edil}  
\left.  
+F_{\mu_1\mu_{2}}^{\, A}  
F^{\mu_1\mu_{2}}_{\,A}  
-2\sqrt{2}F_{\mu_1\mu_{2}}^{\,A}C^{[\mu_1}\partial^{\mu_{2}]}D  
+2C_{[\mu_1}\partial_{\mu_{2}]}D\,C^{[\mu_1}\partial^{\mu_{2}]}D\right]  
\eeq  
for the dilaton;  
\beq  
\label{esca}  
\frac{2}{\sqrt{-g}}\partial_{\mu}\left(\sqrt{-g}\partial^\mu \eta_a\right)=  
\frac{e^{\varphi}}{4}\prod_a e^{\eta_a} F_{\mu_1\mu_{2}}  
F^{\mu_1\mu_{2}}  
-\prod_a e^{-\eta_a}\partial_\mu D \partial^\mu D   
\eeq  
for the scalar fields;  
\beq  
\label{erau}  
\frac{1}{\sqrt{-g}}\partial_{\nu}\left(\sqrt{-g}e^{\varphi}  
\prod_a e^{\eta_a} F^{\nu\mu}\right)=  
e^{\varphi}\partial_\nu D  
\left[  
2 C^{[\mu}\partial^{\nu]}D  
-\sqrt{2}F^{\mu\nu}_{\,A} 
\right]  
\eeq  
for  R-R untwisted  field;  
\beq  
\label{erat}  
\partial_{\nu}\left[\sqrt{-g}e^{\varphi}  
F^{\nu\mu}_{\,A}-\sqrt{2}e^{\varphi}   
C^{[\nu}\partial^{\mu]}D\right]=0  
\eeq  
for R-R twisted  field;  
\beq  
\label{enst}  
\partial_{\mu}\left\{\sqrt{-g}\left[\prod_a e^{-\eta_a}\partial^\mu D  
+e^{\varphi}C_{\nu}\left( C^{[\nu}  
\partial^{\mu]}D-\frac{1}{\sqrt{2}}F^{\nu\mu}_{\,A}   
\right)\right]\right\}=0  
\eeq  
for the NS-NS twisted  field.  
Finally the Einstein equations are:  
\[  
\left(R^{\mu\nu}-\frac{1}{2}g^{\mu\nu}R\right)-  
\left(\partial^\mu\varphi\partial^\nu\varphi-\frac{1}{2}g^{\mu\nu}  
\partial^\gamma\varphi\partial_\gamma\varphi\right)  
-\sum_a\left(\partial^\mu\eta_a\partial^\nu\eta_a-\frac{1}{2}g^{\mu\nu}  
\partial^\gamma\eta_a\partial_\gamma\eta_a\right)  
\]  
\[  
-\frac{e^{\varphi}}{4}\left\{\prod_a e^{\eta_a}  
\left[2F_{\alpha}^{\mu}F^{\alpha\nu}-  
\frac{g^{\mu\nu}}{2}  
F_{\alpha_1\alpha_{2}}F^{\alpha_1\alpha_{2}}\right]  
+\left[2F_{\alpha}^{\mu\,A}F_{\,A}^{\alpha\nu}-\frac{g^{\mu\nu}}{2}  
F_{\alpha_1\alpha_{2}}^{\,A}F_{\,A}^{\alpha_1\alpha_{2}}\right]  
\right.  
\]  
\[  
\left.  
-2\sqrt{2}\left[2  
F_{\alpha}^{\mu\,A}C^{[\alpha}  
\partial^{\nu]}D  
-\frac{g^{\mu\nu}}{2}F_{\alpha_1\alpha_{2}}^{\,A}  
C^{[\alpha_1}\partial^{\alpha_{2}]}D  
\right]  
+2\left[2C_{[\alpha}  
\partial^{\mu]}DC^{[\alpha}  
\partial^{\nu]}D  
\right. 
\right. 
\]  
\beq  
\label{eein}  
\left. 
\left. 
-\frac{g^{\mu\nu}}{2}C^{[\alpha_1}\partial^{\alpha_{2}]}  
DC_{[\alpha_1}\partial_{\alpha_{2}]}D  
\right]  
\right\}  
-\prod_a e^{-\eta_a}\left(\partial^\mu D\partial^\nu D-\frac{1}{2}g^{\mu\nu}  
\partial^\gamma D\partial_\gamma D\right)  
\eeq  
To obtain the fractional D0-brane solution, we make 
the following {\it ansatz} for the metric  
\beq  
\label{ansa}  
ds^2=-B^2(r)dt^2  
+F^2(r)\left(dr^2+r^2 d\Omega^2\right)  
\eeq  
and take  all the fields to be  function only of $r$.  
Under this assumption, introducing the quantity $\xi \equiv  
\ln B + 3\ln F$ the previous equations become  
\[  
\frac{e^{-\xi}}{r^{4}}\partial_r\left(e^{\xi}r^{4}\partial_r\varphi\right)=  
-B^{-{2}}e^{\varphi}\frac{1}{2}  
\left[\frac{1}{2}(\partial_r A_{0})^2  
\right.  
\]  
\beq  
\label{edil2}  
\left.  
+\sqrt{2}(\partial_r A_{0})C_{0}\partial_r D+(C_0\partial_r D)^2  
+\frac{1}{2}\prod_a e^{\eta_a}(\partial_r C_0)^2\right]  
\eeq  
for the dilaton fields.  
\beq  
\label{esca2}  
\frac{e^{-\xi}}{r^{4}}\partial_r\left(e^{\xi}r^{4}\partial_r\eta_a\right)=  
-\frac{1}{4}B^{-{2}}e^{\varphi}\prod_a e^{\eta_a}  
(\partial_r C_0)^2  
-\frac{1}{2}\prod_a e^{-\eta_a}(\partial_r D)^2  
\eeq  
for the scalar fields  
\beq  
\label{erau2}  
\frac{e^{-\xi}}{r^{4}}\partial_r\left(e^{\xi}B^{-{2}}r^{4}  
e^{\varphi}\prod_a e^{\eta_a}\partial_r C_0\right)=  
e^{\varphi} B^{-2}\partial_r D\left[  
 2 C_0\partial_r D+\sqrt{2}\partial_r A_0\right]  
\eeq  
for the R-R untwisted gauge field,   
\beq  
\label{erat2}  
\partial_r\left[e^{\xi}{r^{4}}B^{-2}  
e^{\varphi}\left(\partial_r A_0 +\sqrt{2} C_0\partial_r D\right)\right]=0  
\eeq  
for the R-R twisted vector field  
\beq  
\label{enst2}  
\partial_r\left\{  
e^{\xi}{r^{4}}\left[2 \prod_a e^{-\eta_a}\partial_r D-  
B^{-2}   
e^{\varphi}C_0\left(\sqrt{2}\partial_r A_0  
+2 C_0^2\partial_r D\right)\right]  
\right\}  
=0  
\eeq  
for the NS-NS twisted scalar field.  
And finally  
\[  
R^r_{\,r}=F^{-2}\left[(\partial_r\varphi)^2
+\sum_a(\partial_r\eta_a)^2+\prod_a e^{-\eta_a}(\partial_r D)^2\right]+  
\]  
\beq  
\label{egra2}  
-\frac{3}{8}(BF)^{-2}   
e^{\varphi}\left\{\prod_a e^{\eta_a}(\partial_r C_0)^2  
+ (\partial_r A_0^I)^2+2\sqrt{2} (\partial_r D)C_0\partial_r A_0^I  
+2(C_0\partial_r D)^2\right\}  
\eeq  
\beq  
\label{egra3}  
R^0_{\,0}=-  
\frac{3}{8}(BF)^{-2} e^{\varphi}\left\{\prod_a e^{\eta_a}(\partial_r C_0)^2  
+(\partial_r A_0)^2+2\sqrt{2} (\partial_r D)C_0\partial_r A_0+ 
2(C_0\partial_r D)^2\right\}  
\eeq  
\beq  
\label{egra4}  
R^\theta_{\,\theta}=  
\frac{1}{8}(BF)^{-2} e^{\varphi}\left\{\prod_a e^{\eta_a}(\partial_r C_0)^2  
+(\partial_r A_0)^2+2\sqrt{2} (\partial_r D)C_0\partial_r A_0  
+2(C_0\partial_r D)^2\right\}  
\eeq  
for the Einstein equations.   
To simplify the set of eqs.(\ref{edil2})-(\ref{egra4})   
it is convenient to   
obtain an equation for the quantity  
$\xi$. This is  achieved by combining the   
Einstein equations for   
the $R^{0}_{0}$ and $R^{\theta}_{\,\theta}$ components, obtaining:  
\beq  
\label{s1}   
{\partial^2_r\xi} +\frac{7{\partial_r\xi}}{r} + {(\partial_r\xi )}^2 =0  
\eeq   
As asymptotically $\xi =0$ we take this to be the solution everywhere.   
Then by using the dilaton equation and the one for the $R^{r}_{\,r}$ 
component,  
taking into accounts the asymptotic behavior, one gets:  
\beq  
\label{s2}  
B= F^{-{3}} =e^{-\frac{3}{2}\varphi}   
\eeq  
Expanding the fields up to the second    
order around their asymptotic values,    
one sees that the twisted fields   
do not get corrections with respect to their harmonic asymptotic behavior   
Taking this as an {\it ansatz} for the full solution, one can 
easily solve the field equations. The solution is written  
in section {\bf 4}.

\end{document}